# Negative chemical pressure effects induced by Y substitution for Ca on the 'exotic' magnetic behavior of spin-chain compound, $Ca_3Co_2O_6$

S. Rayaprol and E.V. Sampathkumaran[+]

*Tata Institute of Fundamental Research, Homi Bhabha Road, Colaba, Mumbai-400005*

[+]**Corresponding Author**
Prof. E. V. Sampathkumaran
Department of Condensed Matter Physics & Material Science
Tata Institute of Fundamental Research, Colaba
Mumbai – 400 005 (INDIA)
Telephone # +91-22-22804545
Fax # +91-22-22804610
Email: sampath@tifr.res.in

**Running title:**
Negative chemical pressure effects on $Ca_3Co_2O_6$

**Abstract**

The magnetic behavior of a solid solution, $Ca_{3-x}Y_xCo_2O_6$, based on an 'exotic' spin-chain compound, $Ca_3Co_2O_6$, crystallizing in $K_4CdCl_6$-derived rhombohedral structure is investigated. Among the compositions investigated (x= 0.0, 0.3, 0.5, 0.75 and 1.0), single phase formation persists up to x= 0.75, with the elongation of the *c*-axis. The present investigations reveal that the temperature at which the 'so-called' 'partially disordered antiferromagnetic structure' sets in (which occurs at 24 K for the parent compound, x= 0.0) undergoes gradual reduction with the substitution of Y for Ca, attaining the value of about 2.2 K for the nominal x = 1.0. The trend observed in this characteristic temperature is opposite to that reported under external pressure, thereby establishing that Y substitution exerts negative chemical pressure. Anomalous steps observed in the isothermal magnetization at very low temperatures (around 2 K) for x= 0.0, which have been proposed to arise from 'quantum tunneling effects' are found to vanish by a small substitution (x = 0.3) of Y for Ca. Systematics in ac and dc magnetic susceptibility behavior with Y substitution for Ca have also been probed. We believe that the present results involving the expansion of chain length without disrupting the magnetic chain may be useful to the overall understanding of the novel magnetism of the parent compound.

**INTRODUCTION**

Among spin-chain systems, the compound, $Ca_3Co_2O_6$ [Ref 1,2], crystallizing in a $K_4CdCl_6$-derived rhombohedral structure, has attracted a lot of attention theoretically as well as experimentally in recent years [see, for instance, references 3-13, and articles cited therein]. This is primarily associated with the intrinsic frustration of the magnetic coupling arising from the triangular arrangement of antiferromagnetically coupled ferromagnetic chains. Thus, this compound provides an unique opportunity to probe topological frustration effects among quasi-one-dimensional spin systems. In this compound, there are at least two magnetic transitions (T1 = 24 and T2 = 10 K). While the origin of the 10K- transition has been controversial, available literature viewed together suggests an inhomogeneous nature of the magnetic phase below T2 [6]. The 24K-transition is attributable to a not-so-common 'partially disordered antiferromagnetic (PDA) structure' [3, 4], in which two out of three magnetic chains are antiferromagnetically ordered. Another exciting aspect of this compound is that there are multiple steps in the isothermal magnetization (M) at an interval of about 12 kOe well below 10 K [3,4,7]; there is a recent attempt to explain this in terms of an interesting phenomenon called 'quantum tunneling' [9] usually known among molecular magnets. Other interesting aspects of this compound have been brought out in many recent papers [6-13]. It is therefore of interest to carry out further studies on this 'exotic' magnetic system.

The present work is aimed at understanding the magnetic behavior of this spin-chain system by modifying the interatomic distances along the chain without disrupting the chain. This is achieved by carrying out the substitution of Y for Ca. The present studies reveal that such a substitution essentially expands c-axis, and hence this solid solution serves as an ideal system to probe the role of negative chemical pressure, while the positive pressure effect on T1 and T2 has already been investigated by M studies [13]. We would like to add that, while this work was going on, we came across a recent brief report [14] on the magnetic susceptibility ($\chi$) and transport behavior of this solid solution for x up to 0.5; the present investigation extending to higher x is more exhaustive.

**EXPERIMENTAL DETAILS**

The compounds, $Ca_{3-x}Y_xCo_2O_6$ (x= 0.0, 0.3, 0.5, 0.75 and 1.0), in the polycrystalline form, have been prepared by solid state method as described earlier [7] starting with the high purity (>99.9%) oxides, $CaCO_3$, $Y_2O_3$ and $Co_3O_4$. The x-ray diffractions (Cu $K_\alpha$) are shown in figure 1. Among the compositions taken up for investigation, all are single phase except x= 1.0 in which case few extra lines have been observed (see Fig. 1). The lattice constants are shown in Table 1 and it is clear that Y substitution for Ca causes a remarkable increase in *c,* as evidenced by increasing *c/a* with x. We find that the variation in *a* is negligible as reported in Ref. 14. Dc as well as ac $\chi$ measurements have been performed (1.8 – 300 K) employing a commercial (Quantum Design) superconducting quantum interference device. Isothermal M data were obtained with a vibrating sample magnetometer for the zero-field-cooled (ZFC) conditions of the specimens and the rate of increase of magnetic field (H) was kept the same (3 kOe/min) throughout this investigation. Heat-capacity (C) behavior (1.8 – 40 K) was obtained by a relaxation method with the help of a commercial (Quantum Design) physical property measurements system.

**RESULTS AND DISCUSSION**

The temperature (T) dependencies of dc $\chi^{-1}$ obtained in a magnetic field of 5 kOe are compared for all compositions in Fig. 2. $\chi$ follows Curie-Weiss behavior above 130 K for the parent compound (x = 0.0) and the T-range over which this is observed marginally increases with increasing substitution. The values of the paramagnetic Curie temperature ($\theta_p$) and the effective moment ($\mu_{eff}$) obtained from the high temperature linear region (130 – 300 K) are shown in Table 1. The value of $\theta_p$ is close to 30 K for x= 0.0 and it is clear from the table that $\theta_p$ decreases gradually with increasing Y substitution, attaining a value close to zero for x = 0.75 and a negative value for x = 1.0. This finding implies that Y substitution weakens ferromagnetic coupling in favor of antiferromagnetic interaction and therefore we infer that the expansion along c-axis reduces intrachain interaction which has been known to be ferromagnetic. In contrast to the report of Sekimoto et al [14], $\mu_{eff}$ is essentially constant within the limits of experimental error and therefore we propose that the valency of Co [Ref. 5, 13] at both the trigonal prismatic (trivalent, $3d^6$ high-spin configuration) and the octahedral (trivalent, $3d^6$ low-spin configuration, non-magnetic) sites, are essentially unaltered by Y substitution. This implies that Y substitution does not act an electron donor to Co and the charge imbalance created by the presence of Y may be compensated by possible small changes in oxygen content.

In order to infer the trends in the onset of magnetic ordering, we show the low temperature ZFC $\chi$ behavior obtained at H = 5 kOe and 100 Oe in figure 3. For this purpose, we also look at the temperature at which there is a sudden deviation in the low temperature linear behavior of inverse $\chi(T)$ plots below 30 K, shown in the inset of figure 2, We first look at the 5kOe-data (Fig. 2 inset and Fig. 3). The temperature at which the sudden change occurs due to the onset of the first magnetic transition (T1 = 24 K for x = 0.0) gradually decreases with increasing x (see Table 1 also). This transition occurs around 2.2 K for x = 1.0 (as evidenced by the appearance of a peak in C(T) discussed later in this article). There are peaks and drops at lower temperatures for x = 0.0 below 10 K (T2) in the 5kOe-data as discussed already in Ref. 6, due to the onset of another complex magnetic transition.. This feature also apparently is pushed to lower temperatures gradually with x. The above trends in magnetic transition temperatures are reflected even in the data recorded with H = 100 Oe, though such a low field shows qualitative differences in the nature of the curves due to extreme sensitivity of the magnetism of this sample to magnetic fields.

With respect to ac $\chi$ behavior, it may be recalled [7] (see also Fig. 4) that the parent compound exhibits a large frequency ($\nu$) dependence of the curves below 24 K, which is remarkably suppressed by the application of H of 10 kOe and reappears around 10 K (however with a reduced peak value) in the presence of H= 50 kOe. Such a complex ac $\chi$ behavior also brings out exotic nature of the magnetism of this compound. We have investigated how such features respond with the expansion of c-axis by Y-substitution by measuring ac $\chi$ at four frequencies with an ac amplitude of 1 Oe in the absence and in the presence of dc magnetic fields of 10 and 50 kOe. The imaginary component is noisy in many cases and hence we stick our discussions here to real part only; also, in the presence of high fields, the real part for large ν also are noisy, particularly for x>0.3, and hence we do not show these data. The main observations are: (i) In the absence of an external dc magnetic field, we observe significant ν dependence of ac χ curves for all Y substituted compositions as in the case of the parent compound in the magnetically ordered state. (ii) The peak intensity gradually decreases with x, which implies that the fraction of magnetic ions involved in the complex spin dynamics interestingly decreases with increasing x, the origin of which is unclear to us at present. (iii) For x = 0.3, in the presence of dc H of 10 and 50 kOe, the features are very prominent with a noticeable ν-dependence, though the intensity of the 10K-peak is reduced; this situation is somewhat different from that observed for the parent compound. (iv) The application of H tends to suppress the ac χ features gradually with increasing x with the features completely disappearing for H= 50 kOe for x= 0.75 and 1.0. All these results establish that the spin dynamics undergoes profound changes despite the fact that a non-magnetic chemical substitution takes place only at the Ca site, but not along the magnetic chain.

We now discuss isothermal M behavior. It may be recalled [4, 9] that there are multiple steps in M-H plots *at equal field intervals* (12 kOe) well below T2 in the single crystals of $Ca_3Co_2O_6$, resembling the behavior of molecular-based magnets, and hence a novel phenomenon of 'quantum tunneling of magnetization' was proposed to occur in this compound. This feature can be seen [7] even in polycrystals while increasing or decreasing H (see 1.8 and 5 K data of the parent compound in figure 5). In order to understand this aspect further, it is important to investigate how this feature responds to various perturbations. The present chemical substitution allows us to understand how the c-axis expansion modifies this interesting feature. With this primary motivation, we have obtained isothermal M data at selected temperatures for all compositions till 120 kOe. It is distinctly clear from the figure 5 that, for a small substitution of Y for Ca (that is, x = 0.3), multiple steps are apparently absent at 1.8 and 5 K (that is, below T2), and there is a change of slope around 35 kOe followed by a tendency for saturation at high fields as though there is a broad spin reorientation transition; interestingly, a prominent step appears in the range 10 to 30 kOe at 5 and 8 K, however without any evidence for multiple steps (both for increasing and decreasing field cycle). These results seem to indicate that the weakening of the exchange interaction by elongation of the c-axis profoundly influences the so-called 'quantum tunneling' behavior, though the role played by disorder created by chemical substitution is not clear at present. M-H curves are hysteretic for x= 0.3 as for x= 0.0. For x= 0.5, we have taken the data at 1.8 K, which is below T2. As in the case of x= 0.3, M monotonically increases with H till high fields, however exhibiting hysteretic behavior and a weak spin reorientation around 30 kOe – qualitatively different from that observed for the parent compound in the magnetically ordered state. For x = 0.75 and 1.0 also, at 1.8 K, there are no steps in M-H plots, and M varies linearly with H till about 40 kOe and there is a tendency for saturation at very high fields. It is very important to note that M saturates to similar values at high fields for all compositions, which endorses our conclusion (see above) that Y substitution does not change the valency of Co.

In order to render support to the assignment of T1 in all compositions, the behavior of C is informative. It may be recalled [7] that, at T1 in zero field, one should observe a peak in C(T) due to the onset of long range ordering from 2/3 of the total number of magnetic chains, whereas at T2, no such peak would be observed due to the complex (spin-glass-like) nature of the magnetic phase [6]. We use this criterion to establish our conclusion. From figure 6, it is obvious that a distinct shoulder or a peak is observable at the onset of magnetic ordering for all compositions, including 1.0. We therefore conclude that the ones seen for $x > 0.3$ in C(T) plots represents T1 and the features seen in figures 2 and 3 have been interpreted consistently to tabulate T1 and T2. The actual Y composition for the nominal $x = 1.0$ could be in between 0.75 and 1.0, judged by the trends in the values of T1 as determined from the features in C.

**CONCLUSIONS**

To summarise, we are able to form single phase solid solutions, $Ca_{3-x}Y_xCo_2O_6$, up to 0.75. The Y substitution causes expansion of c-axis (and hence unit-cell volume) and we find that this lattice expansion depresses T1 and T2. Consistent with this observation, the opposite trend has been reported in the external high pressure experiments [12]. Our magnetization data reveal that there is no change in the valency of Co by the substitution of Y for Ca, contradicting the conclusions of Sekimoto et al [14]. The spin dynamics apparently undergoes profound changes as inferred from the ac $\chi$ behavior and also the anomalous magnetization behavior (that is, quantum tunneling behavior) is destroyed by a small Y substitution, despite the fact that this chemical substitution does not take place along the magnetic chain. Possible weakening of the intrachain magnetic coupling caused by c-axis expansion somehow brings about these dramatic changes in the magnetic properties and we hope this inference could contribute to the overall understanding of the magnetism of $Ca_3Co_2O_6$. Finally, we would to mention that all the compositions are found to be insulators.

We thank Kartik K Iyer for his help during measurements.

# REFERENCES

[1] H. Fjellvåg, E. Gulbrandsen, S. Aasland, A. Olsen, B. C. Hauback, J. Solid State Chem. **124,** 190 (1996)

[2] S. Aasland, H. Fjellvåg and B. Hauback, Solid State Commun. **101,** 187 (1997)

[3] H. Kageyama, K. Yoshimura, K. Kosuge, H. Mitamura and T. Goto, J. Phys. Soc. Jpn, **66,** 1607 (1997)

[4] A. Maignan, C. Michel, A. C. Masset, C. Martin and B. Raveau, Eur. Phys. J. B **15,** 657 (2000)

[5] E. V. Sampathkumaran, N. Fujiwara, S. Rayaprol, P. K. Madhu and Y. Uwatoko, Phys. Rev. B. **70,** 014437 (2004)

[6] E. V. Sampathkumaran, Z. Hiroi, S. Rayaprol and Y. Uwatoko, J. Magn. Mag. Mat. **284,** L7 (2004)

[7] S. Rayaprol, Kausik Sengupta and E. V. Sampathkumaran, Solid State Commun. **128,** 79 (2003)

[8] S. Rayaprol, Kausik Sengupta and E. V. Sampathkumaran, Proc. Indian Acad. Sci., Chem. Sci. **115,** 553 (2003)

[9] A. Maignan, V. Hardy, S. Hébert, M. Drillon, M. R. Lees, O. Petrenko, D. McK. Paul and D. Khomskii, J. Mater. Chem **14**, 1231 (2004)

[10] V. Hardy, M. R. Lees, O. A. Petrenko, D. McK. Paul, D. Flahaut, S. Hébert and A. Maignan, Phys. Rev. B **70**, 064424 (2004)

[11] V. Hardy, D. Flahaut, M. R. Lees and O. A. Petrenko, Phys. Rev. B **70**, 214439 (2004)

[12] T. Goko, N. Nomura, S. Takeshita and J. Arai, J. Magn. Magn. Mater. **272-276,** E633 (2004)

[13] K. Takubo, T. Mizokawa, S. Hirata, J. −Y. Son, A. Fujimori, D. D. Sarma, S. Rayaprol and E. V. Sampathkumaran, Phys. Rev. B (*In press*)

[14] T. Sekimoto, S. Noguchi and T. Ishida, J. Phys. Soc. Jpn **73**, 3217 (2004)

**Table 1**

The lattice constants (*a* and *c*), *c/a*, paramagnetic Curie temperature ($\theta_p$) and the effective moment ($\mu_{eff}$) per formula unit obtained from the high temperature (130-300 K) Curie-Weiss region of $\chi$, and magnetic transition temperatures (inferred from a combined look of $\chi$ and the C data) for the compounds, $Ca_{3-x}Y_xCo_2O_6$.

| x | a (± 0.01 Å) | c (± 0.01 Å) | c/a | $\theta_p$ (± 1 K) | $\mu_{eff}$ / formula unit (± 0.1 $\mu_B$) | T1, T2 (K) |
|---|---|---|---|---|---|---|
| 0.0 | 9.070 | 10.370 | 1.143 | 29 | 5.20 | 24, 10 |
| 0.3 | 9.040 | 10.436 | 1.154 | 17 | 5.35 | 20, 6 |
| 0.5 | 9.023 | 10.489 | 1.162 | 11 | 5.13 | 14, 5 |
| 0.75 | 9.001 | 10.554 | 1.172 | 0 | 4.81 | 8, 3 |
| 1.0 | 8.990 | 10.584 | 1.177 | -8 | 4.99 | 2.2, <2 |

**Figure Captions**

Figure 1     X-ray diffraction (Cu K$_\alpha$) of the solid solution, Ca$_{3-x}$Y$_x$Co$_2$O$_6$. The extra lines for x= 1.0 are marked by asterisks.

Figure 2     Inverse of magnetic susceptibility ($\chi$) as a function of temperature (30 – 300 K) for Ca$_{3-x}$Y$_x$Co$_2$O$_6$, obtained in a field of 5 kOe. A straight line is drawn through the high temperature linear region to highlight the x-dependence of $\theta_p$. The low temperature data (1.8 - 40 K) are shown in the insets and a straight line is drawn through the linear regions in the paramagnetic state to highlight deviations due to the onset of magnetic order.

Figure 3     Magnetic susceptibility ($\chi$) behavior below 30 K for Ca$_{3-x}$Y$_x$Co$_2$O$_6$, obtained in a field of 100 Oe and 5 kOe for the zero-field-cooled conditions of the specimens.

Figure 4     Real part of ac susceptibility ($\chi'$) below 30 K obtained in (a) 0, 10 and 50 kOe for x= 0.0 and 0.3 and (b) zero field for x= 0.5, 0.75 and 1.0, for the solid solution, Ca$_{3-x}$Y$_x$Co$_2$O$_6$, at various frequencies.

Figure 5     Isothermal magnetization behavior recorded at selected temperatures for the compounds of the series, Ca$_{3-x}$Y$_x$Co$_2$O$_6$. The thin arrows mark the direction in which the field was varied. Thick vertical arrows mark the steps discussed in the text.

Figure 6     Heat capacity (in zero field) as a function of temperature for Ca$_{3-x}$Y$_x$Co$_2$O$_6$.

Figure 1

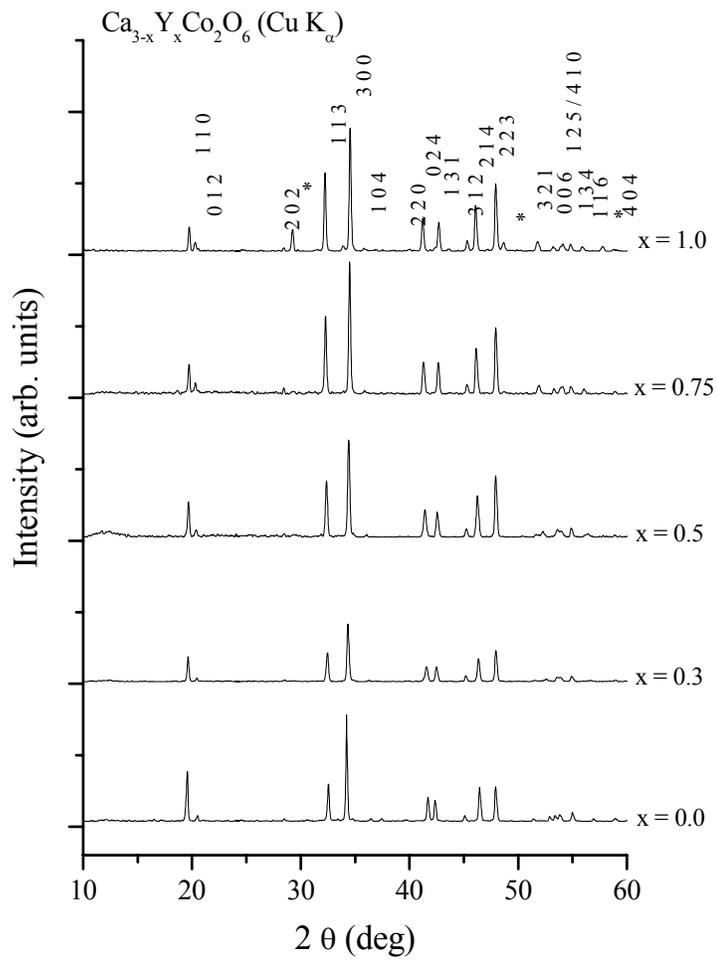

Figure 2

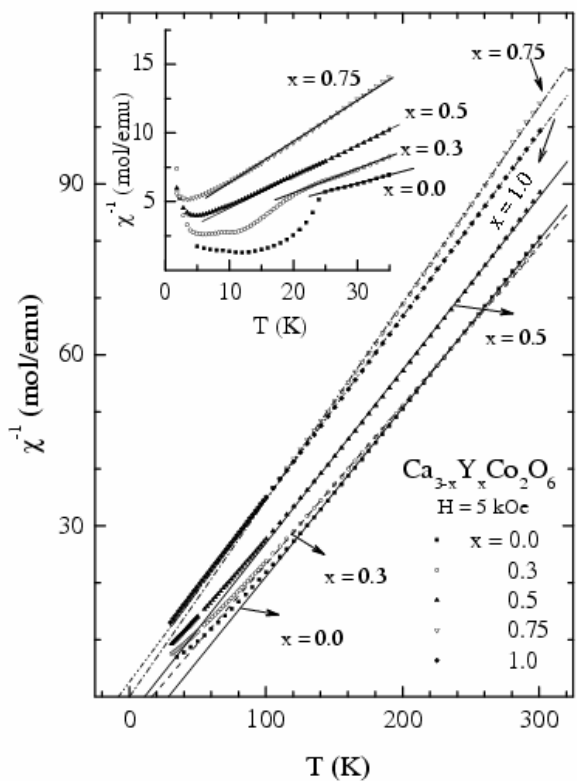

Figure 3

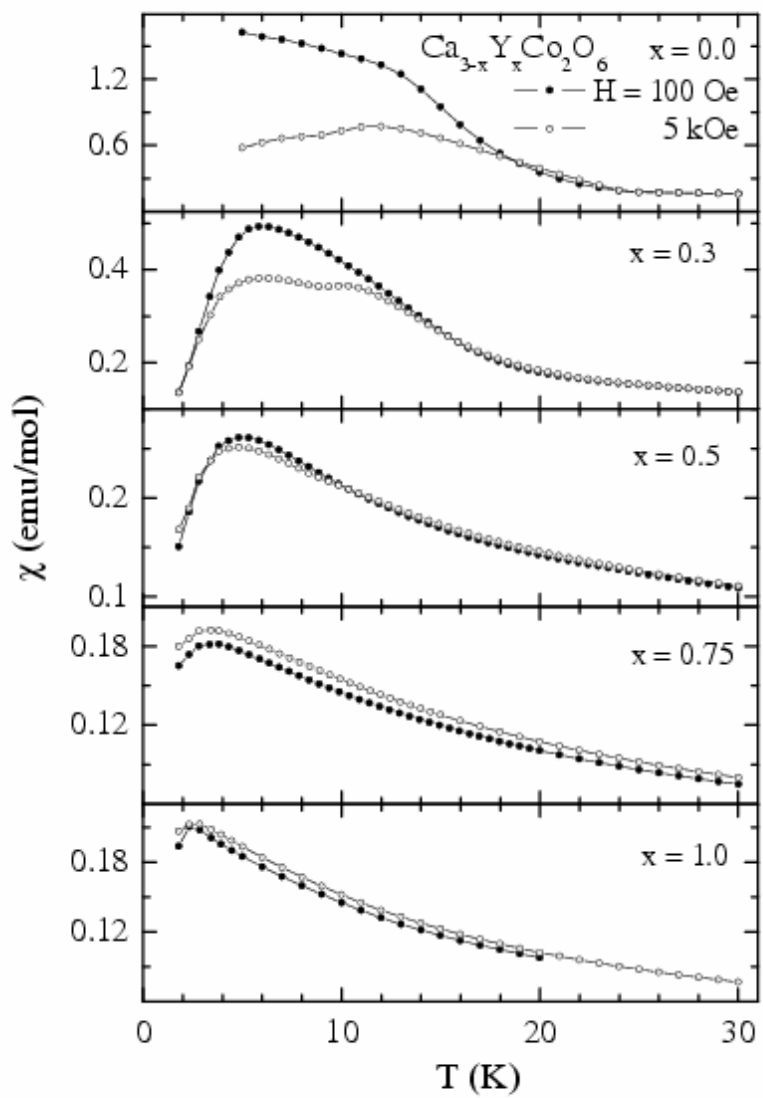

Figure 4

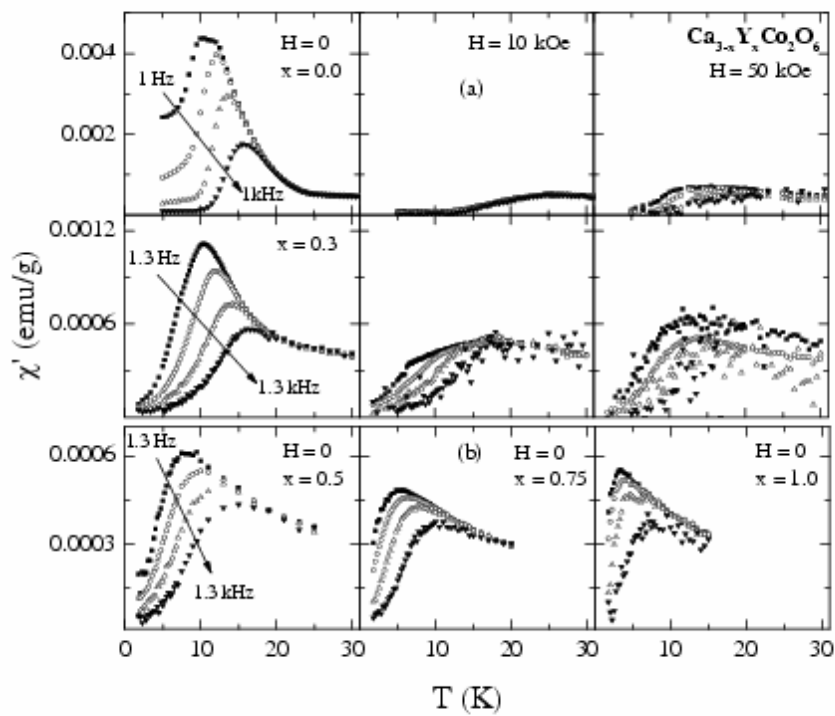

Figure 5

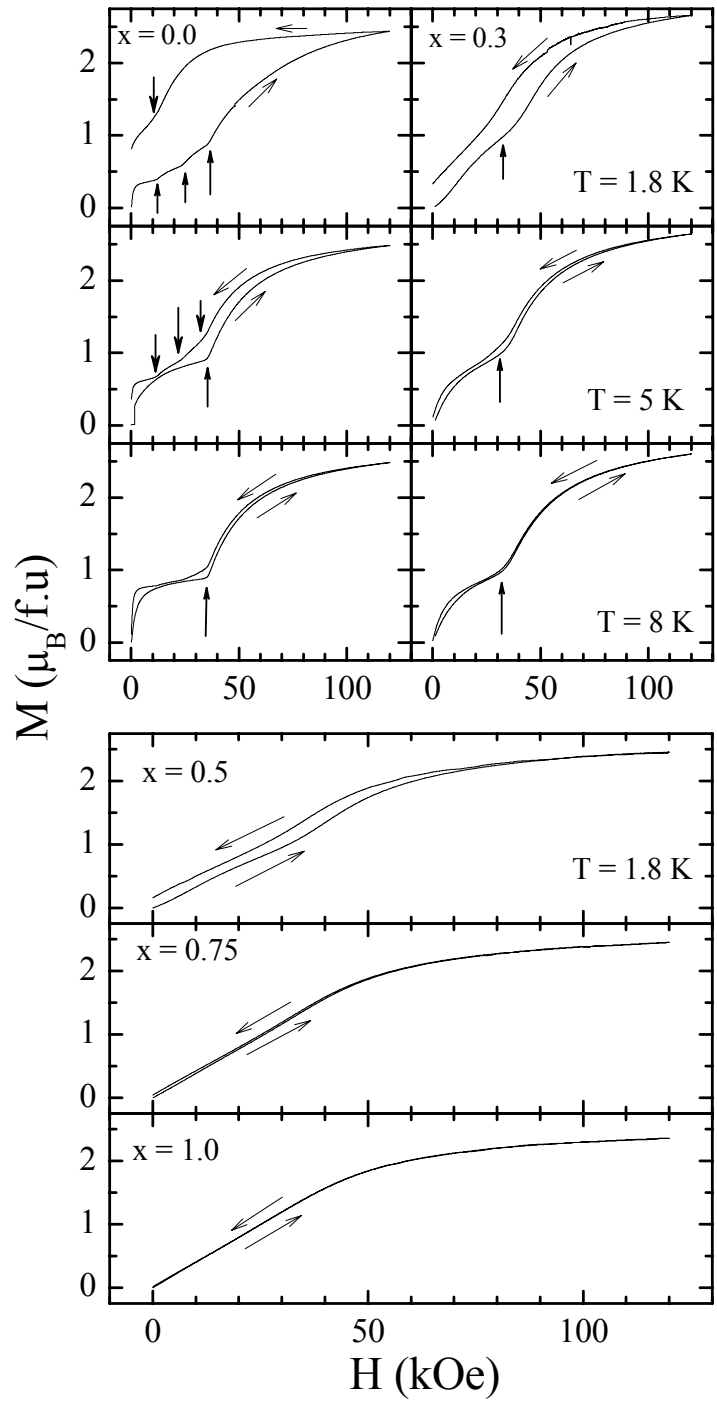

Figure 6

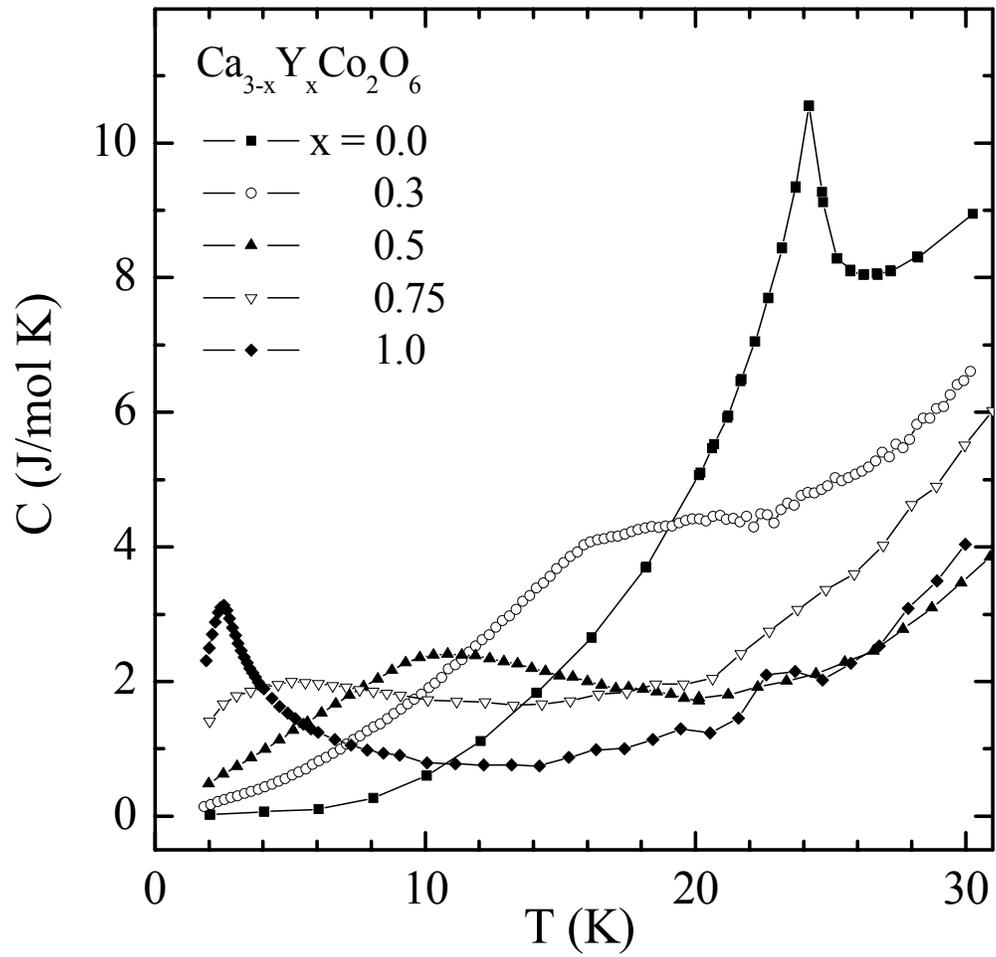